\documentclass{iau} 
\usepackage{graphicx,hyperref}

\title[IAUS334.~~Chemical tagging experiment with the \emph{Gaia}-ESO open clusters] 
{Chemical tagging experiment with the \emph{Gaia}-ESO open clusters}

\author[R.~Smiljanic]   
{Rodolfo Smiljanic$^1$
\and the \emph{Gaia}-ESO Survey consortium}

\affiliation{$^1$Nicolaus Copernicus Astronomical Center, Polish Academy of Sciences, \\ Bartycka 18, 00-716, Warsaw, Poland \\ email: {\tt rsmiljanic@camk.edu.pl}}

\pubyear{2017}
\volume{334}  
\jname{Rediscovering our Galaxy}
\editors{C.~Chiappini, I.~Minchev, E.~Starkenburg \& M.~Valentini, eds.}
\begin{document}

\maketitle

\begin{abstract}
Stars observed in the field of an open cluster are ideal for a controlled test of chemical tagging. Using chemical tagging, one should identify the cluster members, i.e., those stars of similar chemical composition, if their composition is indeed different from that of all the non-member stars of the field. Moreover, the abundance-based membership can be checked against membership based on radial velocities and proper motions. Here, I report preliminary results of such an experiment using data from the \emph{Gaia}-ESO Survey. Although the three membership criteria usually agree, a few interesting examples of discrepant membership classification have been found. In addition, the mean composition of each open cluster was compared to a sample of 1\,600 \emph{Gaia}-ESO field stars. Some cases of field stars with abundances matching those of the open clusters were identified. This experiment suggests that open clusters do not necessarily have unique abundance patterns that set them apart from all other clusters.
\end{abstract}

\firstsection 
\section{Introduction}

Chemical tagging is a potential way to identify field stars that originate in the same star formation event \cite[(Freeman \& Bland-Hawthorn 2002)]{FBH02}. If stars do not change (most of) their surface chemical abundances during their lifetime, then the chemical information is a lifelong ``tag'' shared among stars of common origin. This chemical ``tag'' would mark the stars of the same cluster, long after they have been scattered throughout the Galaxy. There are, however, assumptions that need to be satisfied for chemical tagging to work.

One assumption is that each original stellar cluster was chemically homogeneous. Another is that each original cluster had its own unique chemical signature. If the first assumption is not satisfied, it becomes necessary to connect stars with different abundance patterns to reconstruct a cluster. If the second is not satisfied, chemistry alone is no longer enough to tag stars of common origin.

Both assumptions still need to be validated through observations. For example, a recent high-precision analysis of stars in the Hyades has suggested that at some level clusters might be chemically inhomogeneous \cite[(Liu et al. 2016)]{Liu16}. Moreover, it is important to notice that there is a degree of homogeneity in the trends of abundance ratios with metallicity for most elements \cite[(e.g., Bensby et al. 2014)]{B14}. At a given metallicity, the freedom to have different abundance patterns is perhaps not large enough to guarantee that every single cluster would have its unique chemical composition. 



Regarding the chemical pattern of open clusters, conflicting results exist. While \cite[Blanco-Cuaresma et al. (2015)]{BC15} have found a high degree of overlap in the chemical signatures of open clusters, \cite[Lambert \& Reddy (2016)]{LR16} suggest that the abundances of heavy elements might vary from open cluster to open cluster. Clearly, further work in this area is required to clarify whether chemical tagging might indeed work.

\section{Experiment using the \emph{Gaia}-ESO open clusters}

In this work, the sample of open clusters observed by the \emph{Gaia}-ESO Survey \cite[(Gilmore et al. 2012, Randich \& Gilmore 2013)]{G12,RG13} is used for a controlled test of chemical tagging. In the field of each \emph{Gaia}-ESO open cluster, a large number of stars are selected for observation using the expected location of the cluster in the colour-magnitude diagram. Thus, a good number of the observed stars are expected to be members and it is known a priori that the cluster should be the main peak in the distribution of chemical abundances. Chemical tagging can then be used to identify the cluster members. Moreover, the efficiency of chemical tagging can be tested against other ways to define cluster membership. In particular, using radial velocities (RVs) and proper motions (PMs). 

The data set is made of the FGK-type stars part of the fourth \emph{Gaia}-ESO internal data release (iDR4). Only the results of the analysis of high-resolution UVES spectra were used. Abundances of up to 23 different chemical species are available. The UVES spectra have been analyzed using the \emph{Gaia}-ESO multiple pipelines strategy \cite[(Smiljanic et al. 2014)]{Sm14} with best values estimated as will be described in Casey et al. (in prep). 

Proper-motion based membership probabilities are taken from \cite[Dias et al. (2014)]{Dias14}. The cluster typical RV is taken to be the maximum in the distribution of RVs among the observed stars. Those stars with velocities within $\pm$ 2 km s$^{-1}$ of the maximum are taken to be likely cluster members, those outside $\pm$ 2 km s$^{-1}$ but still within $\pm$ 5 km s$^{-1}$ are taken to be potential members, while those outside a range of $\pm$ 5 km s$^{-1}$ are likely non members of the cluster. This RV criteria was designed to avoid being too restrictive for a first comparison with the results of chemical tagging. 

The chemical tagging itself is implemented using hierarchical clustering, a data mining method of grouping objects based on their similarity. Each star is represented as a point in a multi-dimensional space where the coordinates are the chemical abundances. The distances between the stars in this chemical space are then used to group the similar objects. At the zeroth level, each star is its own group. At successive levels, those points closer together in the chemical space are joined to form a group. The algorithm continues until at the last level, all stars are joined together in one single group.

The group that represents the observed open cluster is searched from the top level to the bottom. At a given level, it is tested if the group with more stars have abundances that all agree within two times the typical abundance error (assumed to be $\pm$ 0.10 dex in [Element/H]). If such a group is not found, the test is repeated at the next level, where the observations are separated in more groups, until a group satisfying this criterium is found. The stars of that group are taken to be the chemical members of the open cluster.

\section{The example of NGC 2243}

\begin{figure}[t]
\begin{center}
\includegraphics[height=3in]{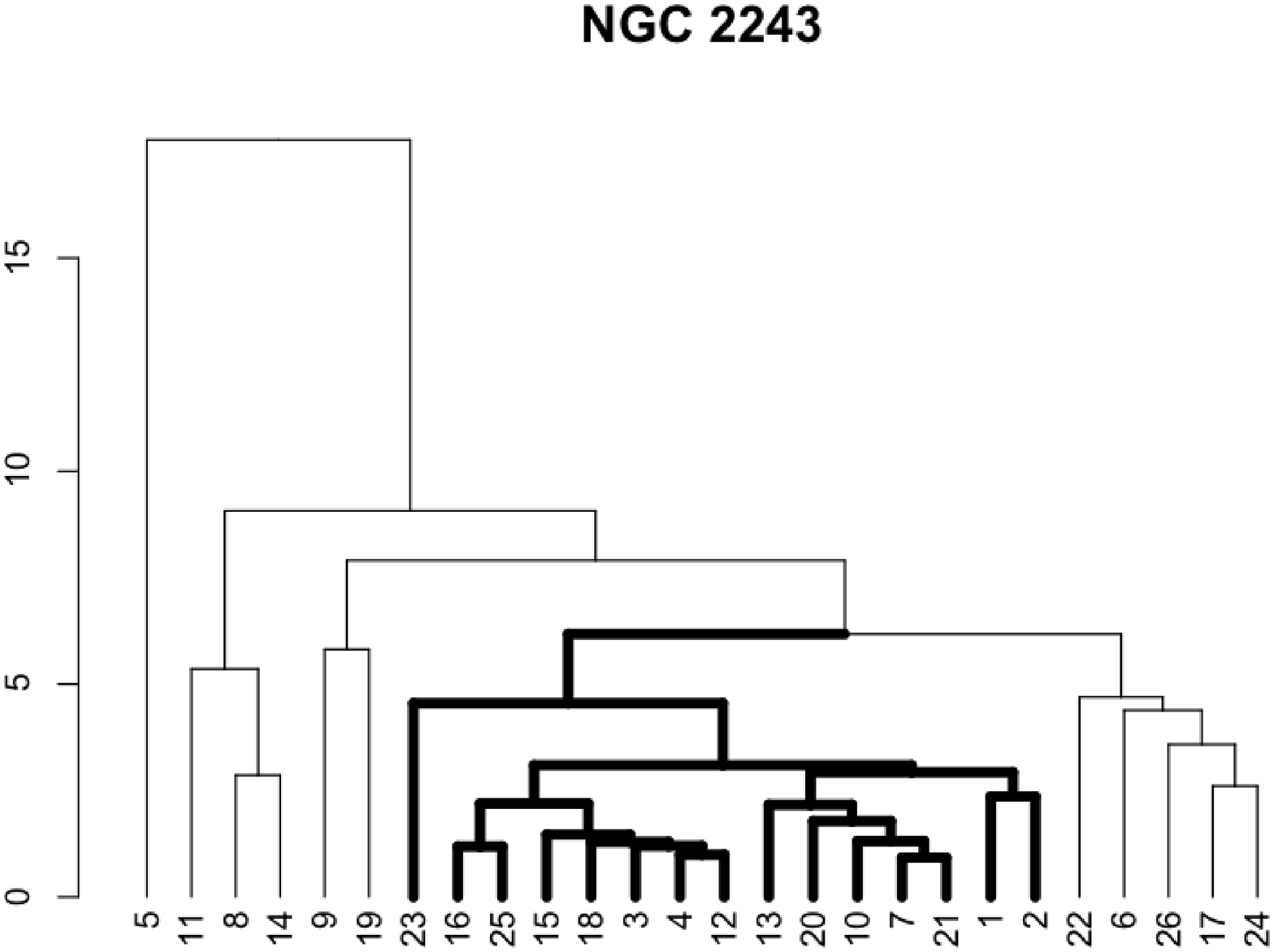}
 \caption{Dendrogram grouping the stars observed in the field of NGC 2243 according to the distance among the objects in the chemical space. The group chosen as the open cluster is highlighted. The numbering is the internal order of the stars in the table used by the algorithm.}
   \label{dend:n2243}
\end{center}
\end{figure}

The results for the metal-poor open cluster NGC 2243 are presented as an example. Abundances for 19 different species were available for a sample of 26 stars. Using chemical tagging, a group of 15 stars was found to be the open cluster (mean abundances in Table \ref{tab:n2243}). The agreement among the abundances of these 15 stars is excellent. The standard deviations of the mean values are below 0.05 dex for most elements. The results of the hierarchical clustering are shown as a dendrogram in Fig. \ref{dend:n2243}.

%
\begin{table}
\begin{center}
\caption{Mean abundance values, and standard deviation (SD), of the chemical abundances in the group of 15 stars found to be the chemical members of NGC 2243.}
\label{tab:n2243}
{\scriptsize
\begin{tabular}{ccccccccccc}
\hline
 & [Fe/H]    & [Al1/Fe] & [Ca1/Fe] & [Ce2/Fe]  & [Co1/Fe] &  [Cr1/Fe] &  [Cr2/Fe] & [Eu2/Fe] & [Mg1/Fe]  & [Mn1/Fe] \\
Mean & $-$0.38  & 0.13  &    0.09   &   0.17  &  $-$0.01 &   $-$0.04   &  0.11  &   0.18   &  0.08 & $-$0.07 \\
SD &     0.04  &  0.03  &    0.03   &  0.09   &  0.04   &   0.05   &  0.06   &  0.08  &   0.03 & 0.06\\
     \hline
 & [Na1/Fe] & [Nd2/Fe] & [Ni1/Fe] & [Sc2/Fe] & [Si1/Fe] & [Ti1/Fe] & [Ti2/Fe] &[V1/Fe] & [Y2/Fe] & \\
Mean &   0.16   &   0.19   & $-$0.03   &  0.09   &  0.02  &   0.04   &  0.16 & $-$0.01 & 0.00 & \\
 SD &  0.05   &   0.09   &  0.04   &  0.05   &  0.03  &   0.04   &  0.02 & 0.10 & 0.04 & \\
\hline
\end{tabular}
}
\end{center}
\end{table}

The typical RV of the cluster is found to be +60.2 km s$^{-1}$. A total of 18 stars were classified as likely members based on the RVs, while 8 are non members. Proper-motion membership probabilities are available for 16 of the 26 stars in the sample. The PMs indicate that 11 stars are likely members (with $>$ 90\% probability) while five are likely non members ($<$ 30\%). The membership classification is given in Table \ref{tab:mem}.

All chemical members are also RV members, except for star \#23 (with RV = +71.7 km s$^{-1}$). This star is also a PM member. In Fig. \ref{dend:n2243}, this star was the last that the hierarchical clustering joined to the group that represent the open cluster. Indeed, it seems somewhat more enhanced in heavy elements than what was found for the cluster (star \#23 has [Fe/H] = $-$0.36, [Ce2/Fe] = +0.27, [Nd2/Fe] = +0.40, and [Y2/Fe] = +0.26). Together with the different RV this perhaps suggests that star \#23 is a barium star, i.e., a binary star with surface abundances changed by material accreted from a companion that has gone through the asymptotic giant branch evolution. This highlights a limitation of chemical tagging. Some stars do change their surface chemical abundances when evolving in a binary system. Although in this experiment star \#23 was included among the cluster members, a star with stronger overabundances would be missed.

Another interesting case is that of stars \#13, \#15 and \#16. Although the chemical abundances and RVs suggest that the stars are members of the cluster, the PM membership probabilities are basically zero. The RVs however, agree very well with the cluster mean (+59.1, +61.2, +60.2 km s$^{-1}$, for stars 
 \#13, \#15 and \#16 respectively). The three stars also have the same effective temperature within the errors ($\sim$ 5000 K) but different $\log~g$ (2.5, 2.8, and 3.1 dex). There are two possible explanations; either they are not cluster members, but field stars with chemical composition similar to the one of the cluster or, despite the RV agreement, they are part of a binary (or multiple) system. The binarity would be responsible for the PM difference. It will be possible to decide which is the correct explanation when better proper motions and parallaxes (distances) for these stars become available from \emph{Gaia}.

\begin{table}
 \caption{Comparison between the three membership classifications for the stars observed in the field of NGC 2243.}
 \label{tab:mem}
\begin{center}
{\scriptsize
\begin{tabular}{lccccccccccccccc}
\hline
{\bf Star Number}    &  1    &     2 &    3  &    4 &    7 &   10 &  12  &  13  & 15   & 16    & 18  & 20 & 21   & 23 & 25 \\
{\bf Radial Velocity} & Yes & Yes & Yes & Yes & Yes & Yes & Yes & Yes & Yes & Yes & Yes & Yes & Yes & Not & Yes \\
{\bf Proper Motions} &   -- &   --  &   -- &    --  & 98\% & 95\% & 97\% & 0\% & 0\% & 2\% & 98\% & 98\% & 95\% & 95\% & -- \\  
{\bf Abundances}    & Yes & Yes & Yes & Yes & Yes & Yes & Yes & Yes & Yes & Yes & Yes & Yes & Yes & Yes & Yes \\
\hline
{\bf Star Number}               &  5 & 6 & 8 & 9 & 11 & 14 & 17 & 19 & 22 & 24 & 26 & & & & \\
{\bf Radial Velocity} & Not & Not & Not & Yes & Not & Not & Not & Yes & Yes & Yes & Not & & & & \\
{\bf Proper Motions} & 91\% & 97\% & -- & 97\% & -- & -- & 29\% & 96\% & 0\% & -- & & & & \\
{\bf Abundances}    & Not & Not & Not & Not & Not & Not & Not & Not & Not & Not & Not & & & & \\
\hline
\end{tabular}
}
\end{center}
\end{table}

A final interesting case is that of stars \#9 and \#19. Both RVs and PMs indicate that these stars are members of the cluster but the chemical abundances are in disagreement. While one star has compatible metallicity (\#9 has [Fe/H] = $-$0.40) the other is too metal-rich (\#19 has [Fe/H] = $-$0.28). They also have other abundances in remarked disagreement, like enhancements in V and Cr. However, the spectra of both stars is of low-signal to noise ($\sim$20). It is thus possible that the abundance results are of low quality. Better data is needed to test the chemical membership status of the two stars.

\section{Comparison with field stars}

In this part of the experiment, the mean cluster abundances were compared to the abundances of a sample of 1\,600 field stars observed by \emph{Gaia}-ESO also with UVES and analyzed in the same way.

Of the eight open clusters analyzed here, field stars with matching abundances are found for seven if a tolerance of $\pm$ 0.14 dex is used. If the search is within $\pm$ 0.10 dex, then matches are still found for five out of the eight clusters. For NGC 2243 in particular, four matching field stars are found in the first case and one in the latter case.

These results, together with the results of the first part of the experiment, seem to show that open clusters do not have unique chemical patterns. If true, this indicates that blind chemical tagging alone would not be sufficient to reconstruct the original clusters that contributed stars to the field.

\begin{acknowledgment}
R. Smiljanic acknowledges support from NCN (2014/15/B/ST9/03981) and from the Polish Ministry of Science and Higher Education.
\end{acknowledgment}

\end{document}